\documentclass[12pt]{article}

\usepackage{color}
\usepackage{a4}
\usepackage{latexsym}
\usepackage{cite}
\usepackage{bm}
\usepackage{lineno}

\usepackage{color}
\usepackage{xcolor}

\usepackage{array}
\newcolumntype{T}{>{\ttfamily} c}
\newcolumntype{N}{>{$\displaystyle} c <{$}}
\newcolumntype{M}{>{$\displaystyle} r <{$}}

\usepackage{epsfig}
\usepackage{graphicx}
\usepackage{amsmath}
\usepackage{amssymb}
\usepackage{multirow}
\allowdisplaybreaks

\usepackage{pslatex}
\usepackage[latin1]{inputenc}
\usepackage[T1]{fontenc}

\usepackage{color,colordvi}
\def\colour4colour#1{\Blue{#1}}

\let\oldbibliography\thebibliography
\renewcommand{\thebibliography}[1]{%
  \oldbibliography{#1}%
  \setlength{\itemsep}{1mm}%
  \setlength{\baselineskip}{3.5mm}
  \setlength{\lineskiplimit}{-\maxdimen}
}

\newcommand{\lsim}{\raisebox{-0.7mm}{$\:\:\stackrel{<}{{\scriptstyle
 \sim}}\: $} }

\newcommand{\beq}{\begin{equation}}
\newcommand{\eeq}{\end{equation}}
\newcommand{\bea}{\begin{eqnarray}}
\newcommand{\eea}{\end{eqnarray}}
\newcommand{\nn}{\nonumber}
\newcommand{\MSb}{$\overline{\mbox{MS}}$}
\newcommand{\ra}{\rightarrow}

\newcommand{\als}{\alpha_{\rm s}}
\newcommand{\ars}{a_{\rm s}}

\newcommand{\hspp}{{\hspace{3mm}}}

\def\frct#1#2{\mbox{\small{$\displaystyle\frac{#1}{#2}$}}}

\def\as(#1){{\alpha_{\rm s}^{\,#1}}}
\def\ar(#1){{a_{\rm s}^{\,#1}}}

\def\zr#1{{\zeta_{\:\!#1}^{}}}

\def\B(#1,#2){{\beta_{#1}^{\,#2}}}

\def\col{\color{black}}

\def\nc{{{n_c}}}
\def\ncs{{{n_{c}^{\,2}}}}

\def\ncf{{{n_{c}^{\,4}}}}

\def\ca{{\col{C^{}_A}}}
\def\cas{{\col{C^{\,2}_A}}}
\def\cat{{\col{C^{\,3}_A}}}
\def\caf{{\col{C^{\,4}_A}}}
\def\cf{{\col{C^{}_F}}}
\def\cfs{{\col{C^{\, 2}_F}}}
\def\cft{{\col{C^{\, 3}_F}}}
\def\cff{{\col{C^{\, 4}_F}}}
\def\nf{{\col{n^{}_{\! f}}}}
\def\nfz{{\col{n^{\,0}_{\! f}}}}
\def\nfo{{\col{n^{\,1}_{\! f}}}}
\def\nfs{{\col{n^{\,2}_{\! f}}}}
\def\nft{{\col{n^{\,3}_{\! f}}}}

\def\dfRAnA{{ \col{{d_{\,R}^{\,abcd\!}\,d_{\,A}^{\,abcd\!} / n_a} }}}
\def\dfRRnA{{ \col{{d_{\,R}^{\,abcd\!}\,d_{\,R}^{\,abcd\!} / n_a} }}}
\def\dfRRnC{{ \col{{d_{\,R}^{\,abcd\!}\,d_{\,R}^{\,abcd\!} / n_c} }}}
\def\dfRAnC{{ \col{{d_{\,R}^{\,abcd\!}\,d_{\,A}^{\,abcd\!} / n_c} }}}

\def\dfAAna{{\col{\frct{d_{\,A}^{\,abcd\!}\,d_{\,A}^{\,abcd\!} }{ n_a} }}}
\def\dfRAna{{\col{\frct{d_{\,R}^{\,abcd\!}\,d_{\,A}^{\,abcd\!} }{ n_a} }}}
\def\dfRRna{{\col{\frct{d_{\,R}^{\,abcd\!}\,d_{\,R}^{\,abcd\!} }{ n_a} }}}
\def\dfRRnc{{\col{\frct{d_{\,R}^{\,abcd\!}\,d_{\,R}^{\,abcd\!} }{ n_c} }}}
\def\dfRAnc{{\col{\frct{d_{\,R}^{\,abcd\!}\,d_{\,A}^{\,abcd\!} }{ n_c} }}}

\def\xm1{{(1 \! - \! x)}}
\def\xp1{{(1 \! + \! x)}}

\def\Lnt(#1){\ln^{\,#1}(1\!-\!x)}

\def\bfkl1{{\mbox{bfkl}_1}}

\def\pqg0N{p_{\rm qg}^{}}
\def\pgq0N{p_{\rm gq}^{}}

\def\z#1{{\zeta_{#1}}}
\def\S(#1){{{S}_{#1}}}
\def\Ss(#1,#2){{{S}_{#1,#2}}}
\def\Sss(#1,#2,#3){{{S}_{#1,#2,#3}}}
\def\Ssss(#1,#2,#3,#4){{{S}_{#1,#2,#3,#4}}}
\def\Sssss(#1,#2,#3,#4,#5){{{S}_{#1,#2,#3,#4,#5}}}
\def\Ssssss(#1,#2,#3,#4,#5,#6){{{S}_{#1,#2,#3,#4,#5,#6}}}
\def\Sssssss(#1,#2,#3,#4,#5,#6,#7){{{S}_{#1,#2,#3,#4,#5,#6,#7}}}

\def\Sp(#1,#2){{{S}_{#1}^{\,#2}}}

\def\H(#1){{\rm{H}}_{#1}}
\def\Hh(#1,#2){{\rm{H}}_{#1,#2}}
\def\Hhh(#1,#2,#3){{\rm{H}}_{#1,#2,#3}}
\def\Hhhh(#1,#2,#3,#4){{\rm{H}}_{#1,#2,#3,#4}}
\def\Hhhhh(#1,#2,#3,#4,#5){{\rm{H}}_{#1,#2,#3,#4,#5}}
\def\Hhhhhh(#1,#2,#3,#4,#5,#6){{\rm{H}}_{#1,#2,#3,#4,#5,#6}}

\def\ddelta{{\delta}}
\def\Dplus(#1){\mathcal{D}_{#1}}
\def\D(#1){{ D_{#1}}}
\def\Dd(#1,#2){{  D_{#1}^{\,#2}}}

\begin{document}
\setlength{\parskip}{0.2cm}
\setlength{\baselineskip}{0.54cm}

% --------------------------------------------------------------------

\begin{titlepage}
\noindent
DESY-25-138 \hfill December 2025 \\
LTH 1410 
\vspace{0.8cm}
\begin{center}
{\LARGE \bf Additional results on the four-loop\\[2mm]
 flavour-singlet splitting functions in QCD}\\
\vspace{1.7cm}
\large
G.~Falcioni$^{\, a}$, F.~Herzog$^{\, b}$, S. Moch$^{\, c}$,
A. Pelloni$^{\, d}$ and A. Vogt$^{\, e\!,\,c\ast}$\\

\vspace{1.0cm}
\normalsize
{\it $^a$Dipartimento di Fisica, Universit\`{a} di Torino,
  and INFN, Sezione di Torino\\
  Via Pietro Giuria 1, 10125 Torino, Italy}\\
\vspace{5mm}
{\it $^b$Higgs Centre for Theoretical Physics, School of Physics and Astronomy\\
  The University of Edinburgh, Edinburgh EH9 3FD, Scotland, UK}\\
\vspace{5mm}
{\it $^c$II.~Institute for Theoretical Physics, Hamburg University\\
\vspace{0.5mm}
Luruper Chaussee 149, D-22761 Hamburg, Germany}\\
\vspace{4mm}
{\it $^d$Institute for Theoretical Physics, 
  ETH Z\"{u}rich, 8093 Z\"{u}rich, Switzerland} \\
\vspace{4mm}
{\it $^e$Department of Mathematical Sciences, University of Liverpool\\
\vspace{0.5mm}
Liverpool L69 3BX, United Kingdom}\\
\vspace{1.7cm}
{\large \bf Abstract}
\vspace{-0.2cm}
\end{center}
We have extended our previous computations, performed analytically for a 
general gauge group, of the even-$N$ moments $\gamma_{\,\rm ik}^{\,(3)}(N)$ of 
the four-loop flavour-singlet splitting functions $P_{\rm ik}^{\,(3)}(x)$ to
$N=22$. 
The numerical QCD results perfectly agree with all predictions resulting from 
the approximations for $P_{\rm ik}^{\,(3)}(x)$ that we obtained before from the
moments $N\!\leq\!20$ and endpoint constraints, confirming their reliability 
for collider-physics applications.
Due to the additional analytical constraints provided by our new $N=22$ 
results, we are now closing in on determining the all-$N$ forms of all
non-rational ($\zeta$-function) contributions to 
$\gamma_{\,\rm ik}^{\,(3)}(N)$: only the $\nfz \,\zr3$ parts of the 
quark-to-gluon (gq) and the $\nfo \,\zr3$ parts of the gluon-to-gluon (qg) 
cases still need to be completed.
Finally we extend our approximations for all $P_{\rm ik}^{\,(3)}(x)$ to 
$\nf = 6$ light flavours and update those for $P_{\rm gq}^{\,(3)}(x)$ at 
lower $\nf$.

\vspace{1cm}
\noindent
$\ast$ {\it Supported by a Research Award of the Alexander-von-Humboldt
Foundation}
\end{titlepage}

% ---------------------------------------------------------------------

In four previous publications 
\cite{Falcioni:2023luc,Falcioni:2023vqq,Falcioni:2024xyt,Falcioni:2024qpd},
we have presented the even Mellin moments to $N = 20$ of the four-loop
flavour-singlet splitting functions $P_{\,\rm ik}^{\,(3)}(x)$ for the 
scale dependence of the parton distributions of hadrons. 
These functions are required for fully consistent analyses of hard 
scattering processes involving initial-state hadrons in perturbative QCD 
at the next-to-next-to-next-to-leading order (N$^3$LO). 
This order constitutes the present accuracy frontier for benchmark
observables, such as the total cross section for Higgs boson production, 
at the Large Hadron Collider (LHC).

From these ten moments, supplemented by all available endpoint constraints,
we have constructed approximate expressions for $P_{\,\rm ik}^{\,(3)}(x)$
that, if the error estimates in refs.~%
\cite{Falcioni:2023luc,Falcioni:2023vqq,Falcioni:2024xyt,Falcioni:2024qpd}
are sound, should be sufficiently accurate for most phenomenological 
applications at the LHC. 
Furthermore we have obtained exact all-$N$ expressions for certain 
zeta-function contributions to their moments
\beq
\label{eq:Mtrf}
  \gamma_{\,\rm ik}^{\,(3)}(N)
  \;=\; -\int_0^1 \!dx\:x^{\,N-1} P_{\,\rm ik}^{\,(3)}(x)
\; .
\eeq
These all-$N$ expressions by themselves are of no phenomenological 
significance, but they can provide useful information on the structure of 
the complete dependence on $N$ which is, in the present flavour singlet 
case, only known for some (numerically mostly small) leading and 
next-to-leading contributions in the limit of a large number of light 
flavours $\nf$ \cite{Davies:2016jie,Gehrmann:2023cqm,Falcioni:2023tzp},
see also refs.~\cite{Gracey:1996ad,Bennett:1997ch}.

The main remaining uncertainties of all the functions 
$P_{\,\rm ik}^{\,(3)}(x)$ are in the small-$x$ (high-energy) region.
In order to remove these, one would need to know all logarithmically
enhanced terms of the form $x^{\,-1} \ln^{\,\ell\!} x$. 
Without this knowledge, the error estimates at $x \lsim 10^{-3}$ remain 
a delicate point. 
Our approximations in refs.~%
\cite{Falcioni:2023luc,Falcioni:2023vqq,Falcioni:2024xyt,Falcioni:2024qpd}
lead to very accurate predictions of the moments at $N \!\geq\! 22$. 
Hence a computation of the next moment(s) entails checks of those error 
estimates, besides providing important input for the determination of 
additional all-$N$ expressions.

We have therefore undertaken to determine all four N$^3$LO anomalous 
dimensions (\ref{eq:Mtrf}) at $N = 22$. 
These have been computed, again using an efficiency-improved in-house
version of the {\sc Forcer} program \cite{Ruijl:2017cxj} in {\sc Form}
\cite{Vermaseren:2000nd,Kuipers:2012rf,Ruijl:2017dtg},
in exactly the same manner as the previous results at $N \leq 20$; 
there is no need to repeat any details of the theoretical framework and 
the computations here. 
At~least for $\gamma_{\,\rm qg}^{\,(3)}(N)$ and 
$\gamma_{\,\rm gg}^{\,(3)}(N)$, an extension to $N=24$ is not feasible 
with present algorithms and computing resources due to the 
prohibitive size of the intermediate expressions for the hardest diagrams.

Our exact results for $\gamma_{\,\rm ik}^{\,(3)}(N\!=\!22)$ are given in 
app.~\ref{sec:appA} for any compact simple gauge group in terms of
rational numbers and the values $\zr3$, $\zr4$, $\zr5$ of Riemann's
$\zeta$-function in the standard \MSb\ factorization for an expansion in 
powers of $\ars = \als/(4\:\!\pi)$. The numerical values for QCD read
\bea
\gamma_{\,\rm ps}^{\,(3)}(N\!=\!22) & =\!\! & 
    \phantom{-93601.653803 + } 
        0.204844770 \,\nf
  \,+\, 0.684781327 \,\nfs
  \,-\, 0.01961886  \,\nft
\, , \nn \\[-0.5mm]
\gamma_{\,\rm qg}^{\,(3)}(N\!=\!22) & =\!\! & 
    \phantom{-93601.653803 + } 
        28.76421340 \,\nf
  \,-\, 104.4064881 \:\nfs
  \,-\, 2.04024428  \,\nft
\, , \nn \\[-0.5mm]
\gamma_{\,\rm gq}^{\,(3)}(N\!=\!22) & =\!\! & 
  -     959.9575206
  \,+\, 90.51612567 \,\nf
  \,+\, 2.269130110 \,\nfs
  \,+\, 0.19217592  \,\nft
\, , \nn \\[-0.5mm]
\gamma_{\,\rm gg}^{\,(3)}(N\!=\!22) & =\!\! & \phantom{-}
    93601.65380
  \,-\, 27096.89412 \,\nf
  \,+\, 1218.733574 \,\nfs
  \,+\, 26.3290244  \,\nft
\, . \quad
\label{eq:gik22-num}
\eea
The corresponding results for $N = 2,\,4,\, \ldots,\, 20\,$ 
can be found
in  eq.~(14) of ref.~\cite{Falcioni:2023luc},
    eq.~(4)  of ref.~\cite{Falcioni:2023vqq},
    eq.~(9)  of ref.~\cite{Falcioni:2024xyt}
and eq.~(5)  of ref.~\cite{Falcioni:2024qpd}. 
Eq.~(\ref{eq:gik22-num}) agrees, of course, with the results in 
refs.~\cite{Davies:2016jie,Gehrmann:2023cqm,Falcioni:2023tzp}.

\begin{table}[t!]
  \centering
  \renewcommand{\arraystretch}{1.2}
  \begin{tabular}{NMM}
\hline\\[-15pt]
   \gamma_{\rm ik}^{\,(3)}(N\!=\!22) & \mbox{computed~~~} & \mbox{predictions~~~~} \\[1pt]
  \hline\\[-5mm]
    \mbox{ps,~ } \nf=3\;  &     6.24785690    &  ~6.2478570(6)~~   \\[0.5mm]
 \phantom{ps,~ } \nf=4\;  &    10.52027295    &  10.5202730(8)~~   \\[0.5mm]
 \phantom{ps,~ } \nf=5\;  &    15.69139890    &  15.6913988(10)   \\[0.5mm]
\hline\\[-5mm]
    \mbox{qg,~ } \nf=3\;  &  -908.4523485     & ~-908.45234(6)~~   \\[0.5mm]
 \phantom{qg,~ } \nf=4\;  & -1686.0225903     & -1686.02258(8)~~   \\[0.5mm]
 \phantom{qg,~ } \nf=5\;  & -2721.3716712     & -2721.37166(10)   \\[0.5mm]
\hline\\[-5mm]
    \mbox{gq,~ } \nf=3\;  &  -662.7982227     &  -662.7981(3)~~    \\[0.5mm]
 \phantom{gq,~ } \nf=4\;  &  -549.2876772     &  -549.2876(3)~~    \\[0.5mm]
 \phantom{gq,~ } \nf=5\;  &  -426.6266492     &  -426.6266(2)~~    \\[0.5mm]
\hline\\[-5mm]
    \mbox{gg,~ } \nf=3\;  & 23990.4572734     & 23990.457275(20)  \\[0.5mm]
 \phantom{gg,~ } \nf=4\;  &  6398.8720777     & ~6398.872080(20)  \\[0.5mm]
 \phantom{gg,~ } \nf=5\;  & -8123.3493836     & -8123.349380(20)  \\[0.5mm]
  \hline
  \end{tabular}
  \caption{\small \label{tab:N=22}
  The (rounded) exact N$^3$LO flavour-singlet $N\!=\!22$ anomalous dimensions 
  computed for this article, compared to the predictions obtained from the 
  $x$-space approximations based on $N \leq 20$ in refs.~%
  \cite{Falcioni:2023luc,Falcioni:2023vqq,Falcioni:2024xyt,Falcioni:2024qpd},
  where the brackets indicate conservative uncertainties of the last digit(s).
  The penultimate digit of ps, $\nf\!=\!5$ and the fourth digit of gg,
  $\nf\!=\!4$ were misprinted in refs.~\cite{Falcioni:2023luc} and 
  \cite{Falcioni:2024qpd}, respectively.
  }
  \vspace*{-2mm}
\end{table}

The comparison of our new results with the $N\!=\!22$ predictions of refs.~%
\cite{Falcioni:2023luc,Falcioni:2023vqq,Falcioni:2024xyt,Falcioni:2024qpd}
for the physically relevant numbers of flavours $\nf = 3,\,4,\,5$ is presented
in table \ref{tab:N=22}. 
The results are in perfect agreement in all 12 cases, indicating that the
uncertainty bands in refs.~%
\cite{Falcioni:2023luc,Falcioni:2023vqq,Falcioni:2024xyt,Falcioni:2024qpd}
do not underestimate the remaining uncertainties at small $x$-values relevant 
to hard-scattering collider observables.

Unlike the results at $N\!=\!20$, the anomalous dimensions at $N\!=\!22$
include an additional prime factor, 23, in the denominators. 
Their inclusion facilitates the determination, as before via systems of 
Diophantine equations, of additional all-$N$ expressions for the $\zr3$ 
contributions -- recall that the $\zr5$ and $\zr4$ parts are completely known 
already
\cite{Falcioni:2023luc,Falcioni:2023vqq,Falcioni:2024xyt,Falcioni:2024qpd,%
Davies:2017hyl}.
In fact, with the exception of the $\nfz$ parts of the quark-to-gluon (gq) 
case and the $\nfo$ terms of the gluon-to-quark (qg) cases, all $\zr3$ 
contributions to the four-loop flavour-singlet anomalous dimensions
have now been determined.
Here we present our ps, qg and gq results. The gg case, where someone else 
provided a crucial step, will be published elsewhere.

\def\col{\color{blue}}

The $\zr3$ part of the N$^3$LO pure-singlet anomalous dimensions can be 
written as
\bea
\label{eq:gps3z3N}
&& \gamma_{\,\rm ps}^{\,(3)}(N)\big|_{\zr3} \;=\;
%%START
%% L texgps3z3N=
  256\,\* \nf\,\* \dfRRnC \* \big\{
    8/3\,\* \eta - 3\,\* \eta^2
  + \S(1) \,\* ( 64/3\,\*\nu - 70\,\*\eta - 68\,\*\eta^2 - 24\,\*\eta^3 )
%%STOP
   \nn \\[-0.5mm] & & \mbox{\hspp}
%%START
  + \S(-2) \,\* ( - 1/3 - 32/3\,\*\nu + 92/3\,\*\eta + 20\,\*\eta^2 )
  + ( 2\* \Ss(1,1) - \S(2)) \,\* ( 27\,\*\eta + 22\,\*\eta^2 - 32/3\,\*\nu )\,
%%STOP
   \nn \\[0mm] & & \mbox{\hspp}
%%START
  + ( \S(3) + 4\* \Ss(-2,1) - 2\* \S(-3) )\,\* ( 3 - 6\*\eta )
 \big\}
  + 32/3\:\* \cf \*\cas\,\*\nf\,\* \big\{
       - 899/18\,\* \D(-1)
       - 2357/12\,\* \D(0)
%%STOP
   \nn \\[0mm] & & \mbox{\hspp} 
%%START
       + 5183/12\,\* \D(1)
       - 1670/9\,\* \D(2)
       - 2/3\,\* \Dd(-1,2)
       - 28/3\,\* \Dd(0,2)
       + 961/6\,\* \Dd(1,2)
       - 622/3\,\* \Dd(2,2)
       - 16\,\* \Dd(-1,3)
%%STOP
   \nn \\[0mm] & & \mbox{\hspp} 
%%START
       + 44\,\* \Dd(0,3)
       + 79\,\* \Dd(1,3)
       - 40\,\* \Dd(2,3)
       - 96\,\* \Dd(0,4)
       + 84\,\* \Dd(1,4)
     + \S(1) \,\* (
       - 54\,\* \D(-1)
       + 749/2\,\* \D(0)
       - 737/2\,\* \D(1)
%%STOP
   \nn \\[0mm] & & \mbox{\hspp} 
%%START
       + 48\,\* \D(2)
       + 40\,\* \Dd(-1,2)
       - 251/2\,\* \Dd(0,2)
       - 245/2\,\* \Dd(1,2)
       + 52\,\* \Dd(2,2)
       + 51\,\* \Dd(0,3)
       - 15\,\* \Dd(1,3) )
     + \Ss(1,1) \,\* (
       - 64/3\,\* \nu
%%STOP
   \nn \\[0mm] & & \mbox{\hspp} 
%%START
       + 54\,\* \eta
       + 44\,\* \eta^2 )
     + \S(2) \,\* (
         44/3\,\* \nu
       - 36\,\* \eta
       - 28\,\* \eta^2 )
     + \S(-2)/3 \,\* (
       - 1
       + 40\,\* \nu
       - 70\,\* \eta
       - 48\,\* \eta^2 )
%%STOP
   \nn \\[0mm] & & \mbox{\hspp}
%%START
     + ( \S(-3) - 2\,\*\Ss(-2,1) - \S(3)/2 ) \,\* (
         3
       - 6\,\* \eta )
     + 2/9\:\*\ddelta(-2+N)
     \big\}
  + 32/3\:\* \cfs\,\*\ca \*\nf\,\* \big\{ 
       - 742/9\,\* \D(-1)
%%STOP
   \nn \\[0mm] & & \mbox{\hspp} 
%%START
       - 1505/4\,\* \D(0)
       + 731/4\,\* \D(1)
       + 4967/18\,\* \D(2)
       + 100/3\,\* \Dd(-1,2)
       + 1395/2\,\* \Dd(0,2)
       - 9/2\,\* \Dd(1,2)
       + 336\,\* \Dd(2,2)
%%STOP
   \nn \\[0mm] & & \mbox{\hspp} 
%%START
       - 335\,\* \Dd(0,3)
       - 49\,\* \Dd(1,3)
       + 56\,\* \Dd(2,3)
       + 282\,\* \Dd(0,4)
       - 186\,\* \Dd(1,4)
     + \S(1) \,\* (
       - 130/3\,\* \D(-1)
       - 323\,\* \D(0)
       + 353\,\* \D(1)
%%STOP
   \nn \\[0mm] & & \mbox{\hspp} 
%%START
       + 40/3\,\* \D(2)
       - 8\,\* \Dd(-1,2)
       + 250\,\* \Dd(0,2)
       + 163\,\* \Dd(1,2)
       - 116\,\* \Dd(2,2)
       - 78\,\* \Dd(0,3)
       - 198\,\* \Dd(1,3) )
     + \Ss(1,1) \,\* (
         48\,\* \nu
%%STOP
   \nn \\[0mm] & & \mbox{\hspp} 
%%START
       - 120\,\* \eta
       - 96\,\* \eta^2 )
     + \S(2) \,\* (
         12\,\* \nu
       - 21\,\* \eta
       - 6\,\* \eta^2 )
     + \S(-2) \,\* (
         32\,\* \nu
       - 72\,\* \eta
       - 48\,\* \eta^2 )
     - 7/9\:\* \ddelta(N-2)
     \big\}
%%STOP
   \nn \\[0mm] & & \mbox{\hspp} 
%%START
    + 32/3\:\,\* \cft\,\*\nf\,\* \big\{
         75\,\* \D(-1)
       + 2673/4\,\* \D(0)
       - 2769/4\,\* \D(1)
       - 51\,\* \D(2)
       - 690\,\* \Dd(0,2)
       - 165\,\* \Dd(1,2)
       - 96\,\* \Dd(2,2)
%%STOP
   \nn \\[-0.5mm] & & \mbox{\hspp} 
%%START
       + 300\,\* \Dd(0,3)
       - 18\,\* \Dd(1,3)
       - 186\,\* \Dd(0,4)
       + 102\,\* \Dd(1,4)
     + \S(1) \,\* (
         16 \,\* \D(-1)
       + 204\,\* \D(0)
       - 240\,\* \D(1)
       + 20\,\* \D(2)
       - 156\,\* \Dd(0,2)
%%STOP
   \nn \\[0mm] & & \mbox{\hspp} 
%%START
       - 72\,\* \Dd(1,2)
       + 96\,\* \Dd(2,2)
       + 96\,\* \Dd(0,3)
       + 144\,\* \Dd(1,3) )
      - ( 3 \* \Ss(1,1) + \S(2) + 3\* \S(-2) ) \,\* (
          16\,\* \nu
        - 36\,\* \eta
        - 24\,\* \eta^2 )
%%STOP
   \quad \nn \\[0mm] & & \mbox{\hspp} 
%%START
     + 2/3\:\* \ddelta(N-2) 
     \big\}
  + 32/9\:\* \cf \*\ca \*\nfs\,\* \big\{
       - 38/3\,\* \D(-1)
       + 1125/3\,\* \D(0)
       - 354\,\* \D(1)
       - 25/3\,\* \D(2)
       - 8\,\* \Dd(2,2)
%%STOP
   \nn \\[0mm] & & \mbox{\hspp}
%%START
       + 16\,\* \Dd(-1,2)
       - 156\,\* \Dd(0,2)
       - 189\,\* \Dd(1,2)
       + 66\,\* \Dd(0,3)
       - 126\,\* \Dd(1,3)
       - 1/3\:\* \ddelta(N-2)
     \big\}
  + 32/9\:\* \cfs\,\*\nfs\,\*  \big\{
         12\,\* \D(-1)
%%STOP
   \quad\nn \\[0mm] & & \mbox{\hspp}
%%START
       - 335\,\* \D(0)
       + 287\,\* \D(1)
       + 36\,\* \D(2)
       - 16\,\* \Dd(-1,2)
       + 145\,\* \Dd(0,2)
       + 187\,\* \Dd(1,2)
       + 8\,\* \Dd(2,2)
       - 66\,\* \Dd(0,3)
       + 90\,\* \Dd(1,3)
%%STOP
   \nn \\[0mm] & & \mbox{\hspp}
%%START
     - \S(1) \,\* (
         8\,\* \nu
       - 18\,\* \eta
       - 12\,\* \eta^2 )
    \big\} 
  + 32/9\:\* \cf \*\nft\,\* \big\{
         4\,\* \D(-1)
       + 3\,\* \D(0)
       - 3\,\* \D(1)
       - 4\,\* \D(2)
       - 6\,\* \Dd(0,2)
       - 6\,\* \Dd(1,2)
     \big\}
%%;
%%STOP
\; .
\eea
\def\col{\color{black}}
Here and below we have used the abbreviations 
\beq
\label{eq:abbr1}
  \D(a) \,=\, (N+a)^{-1} \;,\quad 
  \eta  \,=\, \D(0)  - \D(-1) \quad \mbox{and} \quad
  \nu   \,=\, \D(-1) - \D(2) \; ,
\eeq
and suppressed the argument $N$ of the harmonic sums
$S_{\vec{w}}(N)$ \cite{Vermaseren:1998uu,Blumlein:1998if}.
The $\nft$ and $\nfs$ parts at the end of eq.~(\ref{eq:gps3z3N}) are part
of the complete results in refs.~\cite{Davies:2016jie,Gehrmann:2023cqm}.

Eq.~(\ref{eq:gps3z3N}) includes structures that did not occur in the 
pure-singlet anomalous dimensions to three loops \cite{Vogt:2004mw}.
The first is the presence of terms without any factors $1/(N\!+\!a)$. 
In the present four-loop case, a constant-$N$ contribution was found to
contribute to the simpler $\zr5$ part of $\gamma_{\,\rm ps}^{\,(3)}(N)$
already in ref.~\cite{Moch:2018wjh}; the complete expression was given 
in eq.~(22) of ref.~\cite{Falcioni:2023luc}.
Together with those results, the present $\zr3$-coefficients of 
$S_{-2}(N)$ indicate the presence of the function 
\beq
\label{eq:fNfct}
     f(N) \;=\; 
         5 \,\* \zr5
       + 4 \,\* \zr3 \,\* \S(-2)
       - 2 \,\* \S(-5) - 4 \,\* \Ss(-2,-3) + 8 \,\* \Sss(-2,-2,1)
       + 4 \,\* \Ss(3,-2) - 4 \,\* \Ss(4,1) + 2 \,\* \S(5)
\eeq
that first occurred in the three-loop coefficient functions for inclusive
deep-inelastic scattering (DIS)~\cite{Vermaseren:2005qc}.
We thus predict that the complete expression is of the form
\beq
   \gamma_{\,\rm ps}^{\,(3)}(N) \;=\;
   - \,\frct{8}{9}\:\* ( \cf \*\cas\,\*\nf + 24\,\* \dfRRnC ) \: f(N)
   \:+\: \ldots \quad .
\eeq
The second contribution without denominators, which did not occur in any
three-loop result, is a specific combination of weight-3 sums,
$\,2\,\* S_{-3} - 4\,\* S_{-2,1} - S_{3}\,$. This combination vanishes
as $1/N^2$ (modulo $\ln N$) for $ N \ra \infty$, hence it does not point
to a corresponding non-$\zeta$ structure.

The second new feature of $\gamma_{\,\rm ps}^{\,(3)}(N)|_{\zr3}$
are the $\delta(N\!-\!2)$ terms in the 
$ \cf \nf \{\cfs,\,\cf \ca,\,\cas,\, \ca \nf \}$ colour factors. 
Such contributions are absent in the anomalous dimensions to three loops.
In the DIS coefficient functions, however, they occur already at the
second order \cite{vanNeerven:1991nn,Zijlstra:1991qc,Moch:1999eb}.
These terms arise from non-$\zeta$ contributions 
$(N\!-\!2)^{-1}\, S_{-2}(N\!-\!2)$, in $x$-space corresponding to
$\,x^{\:\!-2}\, {\rm H}_{-1,0}(x) = 
 \,x^{\:\!-2}\, [ \,\ln(x) \ln(1+x) + \mbox{Li}_2(-x) ]$
\cite{Remiddi:1999ew,Moch:1999eb}, see also 
refs.~\cite{Gehrmann:2023cqm,Falcioni:2023tzp}.

\def\col{\color{blue}}
The expressions for the off-diagonal anomalous dimensions corresponding
to eq.~(\ref{eq:gps3z3N}) read
\bea
\label{eq:gqg3z3N}
&& \gamma_{\,\rm qg}^{\,(3)}(N)\big|_{\zr3} \;=\;
%%START
%% L texgqg3z3N=
    32/3\:\* ( \cas + 48\,\* \dfRRnA ) \,\*\nfs \,\* \big\{
     ( - 7/12 
       - 3\,\* \eta
       - 8\,\* \S(1)\,\* \eta^2 
       + 8\,\* \S(-2)\,\* \eta
     ) \,\* \pqg0N
%%STOP
   \nn \\[-0.5mm] & & \mbox{\hspp}
%%START
     + ( 4
       - 16\,\* \S(-2)
     ) \,\* \D(0)\,\*\D(1)\,\*\D(2)
    \big\}
  + 32/3\:\* \cas\,\*\nfs\,\* \big\{
         19/3\,\* \D(-1)
       + 10799/24\,\* \D(0)
       - 3983/12\,\* \D(1)
%%STOP
   \nn \\[0.5mm] & & \mbox{\hspp}
%%START
       - 557/4\,\* \D(2)
       + 8\,\* \Dd(-1,2)
       - 836/3\,\* \Dd(0,2)
       - 700/3\,\* \Dd(1,2)
       - 90\,\* \Dd(2,2)
       + 114\,\* \Dd(0,3)
       - 30\,\* \Dd(1,3)
       - 12\,\* \Dd(2,3)
%%STOP
   \nn \\[0.0mm] & & \mbox{\hspp}
%%START
       - 18\,\* (3\,\*\Dd(0,4) - 2\,\*\Dd(1,4))
     + \S(1) \,\* (
       - 10\,\* \D(-1)
       + 544/3\,\* \D(0)
       - 731/3\,\* \D(1)
       + 421/6\,\* \D(2)
       - 74\,\* \Dd(0,2)
%%STOP
   \nn \\[0.0mm] & & \mbox{\hspp}
%%START
       - 36\,\* \Dd(1,2)
       + 24\,\* \Dd(2,2)
       + 36\,\* (\Dd(0,3) - 2\,\*\Dd(1,3)) )
     - ( 4\,\* \Ss(1,1) + \S(2) - 3\,\* \S(-2) ) \,\* \pqg0N 
     + 1/9\:\* \ddelta(N-2)
     \big\}
%%STOP
   \nn \\[0.0mm] & & \mbox{\hspp}
%%START
  + 32/3\:\* \cf \*\ca\,\*\nfs\,\* \big\{
       - 86/3\,\* \D(-1)
       - 11993/12 \,\* \D(0)
       + 1828/3\,\* \D(1)
       + 462\,\* \D(2)
       - 32/3\,\* \Dd(-1,2)
%%STOP
   \nn \\[0.0mm] & & \mbox{\hspp}
%%START
       + 2035/3 \,\*\Dd(0,2)
       + 1901/3 \,\*\Dd(1,2)
       + 515/3 \,\*\Dd(2,2)
       - 271\,\* \Dd(0,3)
       - 84\,\* \Dd(1,3)
       + 20\,\* \Dd(2,3)
       + 72\,\* (2\,\*\Dd(0,4) - \Dd(1,4))
%%STOP
   \nn \\[0.0mm] & & \mbox{\hspp}
%%START
     + \S(1) \,\* (
         68/3 \,\*\D(-1)
       - 2581/6 \,\*\D(0)
       + 1771/3 \,\*\D(1)
       - 383/2 \,\*\D(2)
       + 165\,\* \Dd(0,2)
       + 86\,\* \Dd(1,2)
       - 62\,\* \Dd(2,2)
%%STOP
   \nn \\[0.0mm] & & \mbox{\hspp}
%%START
       - 84\,\* (\Dd(0,3) - 2\,\*\Dd(1,3)) )
     + ( 10\,\* \Ss(1,1) - \S(2) - 4\,\* \S(-2) ) \* \pqg0N
     - \ddelta(N-2) / 3
     \big\}
  + 32/3\:\* \cfs\,\*\nfs\,\* \big\{
         280/9\,\* \D(-1)
%%STOP
   \quad \nn \\[0.0mm] & & \mbox{\hspp}
%%START
       + 1609/3\,\* \D(0)
       - 1525/6\,\* \D(1)
       - 3085/9\,\* \D(2)
       - 2401/6\,\* \Dd(0,2)
       - 1204/3\,\* \Dd(1,2)
       - 316/3\,\* \Dd(2,2)
       + 153\,\* \Dd(0,3)
%%STOP
   \nn \\[0.0mm] & & \mbox{\hspp}
%%START
       + 110\,\* \Dd(1,3)
       - 16\,\* \Dd(2,3)
       - 24\,\* (4\,\*\Dd(0,4) - \Dd(1,4))
     + \S(1) \,\* (
         517/3\,\* \D(0)
       - 32/3\,\* \D(-1)
       - 770/3\,\* \D(1)
       + 322/3\,\* \D(2)
%%STOP
   \nn \\[0.0mm] & & \mbox{\hspp}
%%START
       - 58\,\* \Dd(0,2)
       - 20\,\* \Dd(1,2)
       + 32\,\* \Dd(2,2)
       + 36\,\* (\Dd(0,3) - 2\,\*\Dd(1,3)) )
     - ( 6\,\* \Ss(1,1) - 2\,\* \S(2) - 4\,\* \S(-2) )\,\* \pqg0N
     + 1/9\,\* \ddelta(N-2)
     \big\}
%%STOP
   \nn \\[0.0mm] & & \mbox{\hspp}
%%START
  + 32/9\:\* \ca\,\*\nft\,\* \big\{
         4/3\,\* \D(-1)
       + \D(0)
       + 4\,\* \D(1)
       - 19/3\,\* \D(2)
       - 2\,\* \Dd(0,2)
       - 4\,\* \Dd(1,2)
       - 4\,\* \Dd(2,2)
       - 2\,\* \S(1)\,\* \pqg0N
     \big\}
%%STOP
   \nn \\[0.0mm] & & \mbox{\hspp}
%%START
  + 32/9\:\* \cf\,\*\nft\,\* \big\{
         16/3\,\* \D(-1) 
       - 107/2\,\* \D(0) 
       + 92\,\* \D(1) 
       - 136/3\,\* \D(2) 
       + 17\,\* \Dd(0,2) 
       - 2\,\* \Dd(1,2) 
       - 16\,\* \Dd(2,2) 
%%STOP
   \nn \\[0.0mm] & & \mbox{\hspp}
%%START
       - 12\,\* \Dd(0,3) 
       + 24\,\* \Dd(1,3) 
       + 2\,\* \S(1)\,\* \pqg0N
     \big\}
%%;
%%STOP
   \:\:+\:\: \mbox{$\nfo$ contributions}
\eea
and
\bea
\label{eq:ggq3z3N}
&& \gamma_{\,\rm gq}^{\,(3)}(N)\big|_{\zr3} \;=\;
%%START
%% L texggq3z3N=
    32/3\:\* ( \cf\*\cas + 24\,\*\dfRRnC ) \,\*\nf \,\* \big\{
     ( - 7/12
       - 3\,\* \eta
       - 8\,\* \S(1)\,\* \eta^2
       + 8\,\* \S(-2)\,\* \eta
     ) \,\* \pgq0N
%%STOP
   \nn \\[-0.5mm] & & \mbox{\hspp}
%%START
     + ( 4
       - 16\,\* \S(-2)
     ) \,\* \D(-1)\,\*\D(0)\,\* \D(1)
    \big\}
  + 32/9\:\* \cf\* \cas\,\*\nf\,\* \big\{
         865/4\,\* \D(-1)
       - 13259/12\,\* \D(0)
       + 24671/24\,\* \D(1)
%%STOP
   \nn \\[0.0mm] & & \mbox{\hspp}
%%START
       + 260/3\,\* \D(2)
       + 31\,\* \Dd(-1,2)
       + 803\,\* \Dd(0,2)
       + 308\,\* \Dd(1,2)
       + 76\,\* \Dd(2,2)
       + 36\,\* \Dd(-1,3)
       + 24\,\* \Dd(1,3)
       - 54 \* ( \Dd(0,3) - 2\,\*\Dd(0,4) 
%%STOP
   \quad \nn \\[0.0mm] & & \mbox{\hspp}
%%START
         + \Dd(1,4))
     + \S(1) \,\* (
         40\,\* \D(-1)
       + 122\,\* \D(0)
       - 257/2\,\* \D(1)
       + 6\,\* \D(2)
       - 90\,\* \Dd(-1,2)
       + 30\,\* \Dd(0,2)
       - 111\,\* \Dd(1,2)
       )
     - ( 24\,\* \Ss(1,1) 
%%STOP
   \quad \nn \\[0.0mm] & & \mbox{\hspp}
%%START
       + 18\,\* \S(2) 
       + 27\,\* \S(-2) ) \,\* \pgq0N
    - 2\,\* \ddelta(N-2)
    \big\}
  + 32/9\:\* \cfs\,\*\ca\,\*\nf\,\* \big\{
       + 191/2\,\* \D(-1)
       + 7642/3\,\* \D(0)
%%STOP
   \quad \nn \\[0.0mm] & & \mbox{\hspp}
%%START
       - 32591/12\,\* \D(1)
       - 220/3\,\* \D(2)
       + 74\,\* \Dd(-1,2)
       - 2289\,\* \Dd(0,2)
       - 1373/2\,\* \Dd(1,2)
       - 96\,\* \Dd(2,2)
       + 360\,\* \Dd(0,3)
       + 39\,\* \Dd(1,3)
%%STOP
   \quad \nn \\[0.0mm] & & \mbox{\hspp}
%%START
       - 72\,\* \Dd(-1,3)
       - 288\,\*  (\Dd(0,4) - \Dd(1,4))
     + \S(1) \,\* (
       - 118\,\* \D(-1)
       - 700 \,\*\D(0)
       + 674\,\* \D(1)
       - 28\,\* \D(2)
       + 24\,\* \Dd(-1,2)
%%STOP
   \quad \nn \\[0.5mm] & & \mbox{\hspp}
%%START
       + 240\,\* \Dd(0,2)
       + 348\,\* \Dd(1,2)
       - 72\,\* (2\,\*\Dd(0,3) - \Dd(1,3))
       )
    + ( 102\,\* \Ss(1,1) -  54\,\* \S(2) - 18\,\* \S(-2) ) \,\* \pgq0N
    + 5\,\* \ddelta(N-2)
    \big\}
%%STOP
   \quad \nn \\[0.0mm] & & \mbox{\hspp}
%%START
  + 32/9\:\* \cft\,\*\nf\,\* \big\{
       - 1372/3\,\* \D(-1)
       - 2233/2\,\* \D(0)
       + 6049/4\,\* \D(1)
       - 20/3\,\* \D(2)
       - 16\,\* \Dd(-1,2)
       + 1477\,\* \Dd(0,2)
%%STOP
   \quad \nn \\[0.0mm] & & \mbox{\hspp}
%%START
       + 305\,\* \Dd(1,2)
       + 32\,\* \Dd(2,2)
       - 186 \,\*\Dd(0,3)
       - 87\,\* \Dd(1,3)
       + 216\,\* (\Dd(0,4) - \Dd(1,4))
     + \S(1) \,\* (
       - 30\,\*\D(-1)
       + 856 \,\*\D(0)
       - 725\,\* \D(1)
%%STOP
   \quad \nn \\[0.5mm] & & \mbox{\hspp}
%%START
       + 32\,\* \D(2)
       + 96\,\* \Dd(-1,2)
       - 288\,\* \Dd(0,2)
       - 300\,\* \Dd(1,2)
       + 108\,\* (2\,\*\Dd(0,3) - \Dd(1,3))
       )
    - ( 78\,\* \Ss(1,1) - 72\,\* \S(2) - 60\,\* \S(-2) ) \,\* \pgq0N
%%STOP
   \quad \nn \\[0.5mm] & & \mbox{\hspp}
%%START
    - 10/3\,\* \ddelta(N-2)
    \big\}
  + 64/27\,\* \cfs\,\* \nfs\,\* \big\{
       + 52\,\*\D(-1)
       - 89\,\*\D(0)
       + 76\,\*\D(1)
       + 4\,\*\D(2)
       + 12\,\*\Dd(-1,2)
       + 42\,\*\Dd(0,2)
%%STOP
   \quad \nn \\[0.5mm] & & \mbox{\hspp}
%%START
       + 9\,\*\Dd(1,2)
    - 6\,\* \S(1) \,\*  \pgq0N
    \big\}
  + 32/9\:\* \cf\*\ca\,\*\nfs\,\* \big\{
         63\,\*\D(0)
       - 103/3\,\*\D(-1)
       - 54\,\*\D(1)
       - 8/3\,\*\D(2)
       - 8\,\*\Dd(-1,2)
%%STOP
   \quad \nn \\[0.5mm] & & \mbox{\hspp}
%%START
       - 28\,\*\Dd(0,2)
       - 12\,\*\Dd(1,2)
    + 6\,\* \S(1) \,\*  \pgq0N
    + 1/3\,\* \ddelta(N-2)
    \big\}
  - 128/27\:\* \cf\,\*\nft\,\* \pgq0N
%%;
%%STOP
  \:\:+\:\: \mbox{$\nfz$ contributions}
\eea
with 
\beq
  \pqg0N \;=\; \D(0) - 2\:\! \D(1) + 2\:\! \D(2)
  \;\;, \quad
  \pgq0N  \;=\; 2\:\! \D(-1) - 2\:\! \D(0) + \D(1)
  \; . 
\eeq
\def\col{\color{black}}
Also $\gamma_{\,\rm qg}^{\,(3)}(N)$ and $\gamma_{\,\rm gq}^{\,(3)}(N)$
include non-$\zeta$ terms $(N\!-\!2)^{-1}\, S_{-2}(N\!-\!2)$, as shown by 
the $\delta(N\!-\!2)$ contributions to eqs.~(\ref{eq:gqg3z3N}) and 
(\ref{eq:ggq3z3N}). The coefficients of these non-$\zeta$ terms can be
read off eqs.~(\ref{eq:gps3z3N}), (\ref{eq:gqg3z3N}) and (\ref{eq:ggq3z3N}), 
taking into account that the second moment of 
$\,x^{\:\!-2}\, {\rm H}_{-1,0}(x)$ is $-3/2\:\*\zr3$.
Besides the results given in eqs.~(\ref{eq:gqg3z3N}) and (\ref{eq:ggq3z3N}), 
only the respective $\nf\, \dfRAnA$ and $\dfRAnC$ $\zr3$-terms are presently
known at all $N$, see eq.~(7) of ref.~\cite{Falcioni:2023vqq} and eq.~(14) 
of ref.~\cite{Falcioni:2024xyt}.

\vspace*{2mm}
To summarize, we have extended our previous computations of the four-loop
flavour-singlet anomalous dimensions to $N=22$.
These results have been used to check the predictions of our approximations 
for the N$^3$LO splitting functions $P_{\rm ps}^{\,(3)}(x)$ and 
$P_{\rm qg\,,gq,\,gg}^{\,(3)}(x)$ given in refs.~%
\cite{Falcioni:2023luc,Falcioni:2023vqq,Falcioni:2024xyt,Falcioni:2024qpd}
for $\nf = 3,\,4,\,5$ light flavours.
Our new results are well inside the error bands, reinforcing that 
these approximations can be safely employed in analyses of parton 
distributions and hard-scattering processes.
Additional approximations for $\nf\!=\! 6$ for application at ultra-high
scales have been requested by several groups, they are presented in 
app.~B together with a minor update for $P_{\rm gq}(x)$.

With the help of the new results at $N=22$, most of the $\zeta$-function 
contributions to the flavour-singlet anomalous dimensions are now known at 
all $N$, only the quadratic-Casimir $\nfo$ parts of 
$\gamma_{\rm qg}^{\,(3)}(N)|_\zr3$ and $\nfz$ parts of 
$\gamma_{\rm gq}^{\,(3)}(N)|_\zr3$ are not known yet.
These all-$N$ expressions should be useful in further research towards 
the complete determination of the four-loop splitting functions.

Corresponding extensions of previous computations to (at least) $N=22$ 
have been performed also for the non-singlet anomalous dimensions which 
are already fully known in the limit of a large number of colours $\nc$ 
\cite{Moch:2017uml}.
Exact $N$-space results, and improved approximations of the 
\mbox{large-$\nc$} suppressed parts of the splitting functions 
$P_{\,\rm ns}^{\,(3)}(x)$, will be presented elsewhere \cite{Moch:2025xxx}.
For~recent progress in the non-singlet sector see also refs.~%
\cite{Gehrmann:2023iah,Kniehl:2025jfs,Kniehl:2025ttz}. 
%
% ---------------------------------------------------------------------
%
\vspace*{-2mm}
\subsection*{Acknowledgements}
\vspace*{-3mm}
This work has been supported by the UKRI FLF MR/Y003829/1,
the ERC Consolidator Grant 101169614,
the STFC Consolidated Grants ST/X000494/1 and ST/X000699/1;
the EU Marie Sklodowska-Curie grant 101104792;
the DFG through the Research Unit FOR 2926,
project number 40824754, and DFG grant MO~1801/4-2, 
the ERC Advanced Grant 101095857;
and the Alexander-von-Humboldt Foundation through a Research Award.

{\small
%\addtolength{\baselineskip}{-2.5mm}
\providecommand{\href}[2]{#2}\begingroup\raggedright\endgroup
}
\vspace{-5mm}
\appendix
%
% ---------------------------------------------------------------------
%
\renewcommand{\theequation}{\ref{sec:appA}.\arabic{equation}}
\setcounter{equation}{0}
\section{The exact anomalous dimensions at $\bm N\!=\!22$}
\label{sec:appA}
%
% ---------------------------------------------------------------------

\vspace{-2mm}
The anomalous dimensions $\gamma^{\,(3)}_{\,\rm ik}(N)$ at even 
$2\leq N \leq 20$ have been computed and presented for a general 
compact simple gauge group in refs.~\cite{Falcioni:2023luc,Falcioni:2023vqq,%
Falcioni:2024xyt,Falcioni:2024qpd,Moch:2021qrk,Moch:2023tdj}.
Here we report the corresponding exact expressions for $N = 22$. 
Their rounded numerical values in QCD have been given in 
eq.~(\ref{eq:gik22-num}) above.

The quadratic Casimir invariants in SU$(\nc)$ are $\ca = \nc$ and
$\cf = (\ncs-1)/(2\nc)$.
The quartic group invariants are products of two symmetrized traces of four
generators $T_r^a$,
\beq
\label{eq:d4def}
  d_{r}^{\,abcd} \; =\; \frct{1}{6}\: {\rm Tr} \, ( \,
   T_{r}^{a\,} T_{r}^{b\,} T_{r}^{c\,} T_{r}^{d\,}
   + \,\mbox{ five $bcd$ permutations}\, )
\; ,
\eeq
in the fundamental ($R$) or adjoint ($A$) representation,
which leads to
\beq
\label{eq:quarticsSUN}
\dfAAna \,=\,
  \frct{1}{24}\: \ncs ( \ncs + 36 )
\:, \;\;
  \dfRAna \,=\,
  \frct{1}{48}\: \nc ( \ncs + 6 )
\:, \;\;
\dfRRna \,=\,
\frct{( \ncf - 6\,\ncs + 18 )}{96\,\ncs}
\;\;
\eeq
with $n_a = \ncs - 1$, and thus 
$d_{\,A}^{\,abcd\!}\,d_{\,A}^{\,abcd\!} = 135$, 
$d_{\,R}^{\,abcd\!}\,d_{\,A}^{\,abcd\!} = 15/2$ and 
$d_{\,R}^{\,abcd\!}\,d_{\,R}^{\,abcd\!} = 5/12$ in QCD.

\def\col{\color{blue}}

The (negative) $N=22$ moments (\ref{eq:Mtrf}) of the four-loop (N$^3$LO) 
flavour-singlet splitting functions, in the standard \MSb\ factorization 
scheme for an expansion in powers of $\ars = \als/(4\:\!\pi)$, are given by
{\small
\bea
\label{eq:gns3N22}
 && \gamma_{\,\rm ps}^{\,(3)}(N\!=\!22)\:=
%%START
%%L texgps3N22=
\nf\,\*\dfRRnc\,\*\Big(\frac{4771652488240839222919134900422144941}{76483200416054917115782082131200000}
%%STOP
\nn \\[0.5mm] & & \mbox{\hspp}
%%START
+\frac{173997356719548534957639149}{4779301517318176502553900}\,\*\z3
-\frac{423441280}{4032567}\,\*\z5\Big)
%%STOP
\nn \\[0.5mm] & & \mbox{\hspp}
%%START
+\nf\,\*\cf\,\*\cas\,\*\Big(\frac{1831358367634045627224427424119534549689599672007632087}{831778200084908674750810296804135166281948313651200000}
%%STOP
\nn \\[0.5mm] & & \mbox{\hspp}
%%START
+\frac{288578520276256276947996283139}{121356024127743137752848628800}\,\*\z3
+\frac{37058312318711}{29987898183243}\,\*\z4
-\frac{43252760}{12097701}\,\*\z5\Big)
%%STOP
\nn \\[0.5mm] & & \mbox{\hspp}
%%START
+\nf\,\*\cfs\,\*\ca\,\*\Big(-\frac{9537919946411547709924507049793808769430551963371399901}{1663556400169817349501620593608270332563896627302400000}
%%STOP
\nn \\[0.5mm] & & \mbox{\hspp}
%%START
-\frac{1243658345859994178447}{1058651014306939797744}\,\*\z3
-\frac{53983543516568}{29987898183243}\,\*\z4
+\frac{645160}{4032567}\,\*\z5\Big)
%%STOP
\nn \\[0.5mm] & & \mbox{\hspp}
%%START
+\nf\,\*\cft\,\*\Big(\frac{27144237576677130838711174348368669911076088902408313217}{3327112800339634699003241187216540665127793254604800000}
%%STOP
\nn \\[0.5mm] & & \mbox{\hspp}
%%START
-\frac{318367588091247494265721}{361264658632243205980140}\,\*\z3
+\frac{5641743732619}{9995966061081}\,\*\z4
-\frac{1290320}{1344189}\,\*\z5\Big)
%%STOP
\nn \\[0.5mm] & & \mbox{\hspp}
%%START
+\nfs\,\*\cf\,\*\ca\,\*\Big(\frac{66744534859555440760197159616828746745698539}{186419718408057746254017507776948489273088000}
+\frac{880814272}{5843189583}\,\*\z3
%%STOP
\nn \\[0.5mm] & & \mbox{\hspp}
%%START
-\frac{645160}{4032567}\,\*\z4\Big)
+\nfs\,\*\cfs\,\*\Big(-\frac{5099805882936683284299353755270960005973264367}{22459137503446957048698299746460937040995840000}
%%STOP
\nn \\[0.5mm] & & \mbox{\hspp}
%%START
-\frac{86345188824326}{269891083649187}\,\*\z3
+\frac{645160}{4032567}\,\*\z4\Big)
%%STOP
\nn \\[0.5mm] & & \mbox{\hspp}
%%START
+\nft\,\*\cf\,\*\Big(-\frac{468362287708579371348464949691769}{11605762011432587235855925766659200}
+\frac{258064}{12097701}\,\*\z3\Big)
%%;
%%STOP
\; , 
\\[3mm] 
\label{eq:gqg3N22}
 &&\gamma_{\,\rm qg}^{(\!3)}(N\!=\!22)\:=
%%START
%%L texgqg3N22=
\nf\,\*\dfRAna\,\*\Big(-\frac{3810637698789182589166962695084388793}{50988800277369944743854721420800000}
%%STOP
\nn \\[0.5mm] & & \mbox{\hspp}
%%START
-\frac{6094498287658989563858609839}{22940647283127247212258720}\,\*\z3
+\frac{2229427574560}{5644249611}\,\*\z5\Big)
%%STOP
\nn \\[0.5mm] & & \mbox{\hspp}
%%START
+\nf\,\*\cat\,\*\Big(\frac{17275619268122124906361907340026494259937400517669185527192679}{111764373189009008808916877960978034022972831008684441600000}
%%STOP
\nn \\[0.5mm] & & \mbox{\hspp}
%%START
+\frac{4237301575461501673836836219981}{153291819950833437161493004800}\,\*\z3
-\frac{66060918332857867759}{575630557583713632}\,\*\z4
%%STOP
\nn \\[0.5mm] & & \mbox{\hspp}
%%START
-\frac{4819148746285}{67730995332}\,\*\z5\Big)
%%STOP
\nn \\[0.5mm] & & \mbox{\hspp}
%%START
+\nf\,\*\cf\,\*\cas\,\*\Big(-\frac{1723671160224114092426578739601775966167925924769421470190393667}{3352931195670270264267506338829341020689184930260533248000000}
%%STOP
\quad \nn \\[0.5mm] & & \mbox{\hspp}
%%START
-\frac{649796567719936732898841253461977}{7281361447664588265170917728000}\,\*
\z3
+\frac{12669712860380923416023}{60441208546289931360}\,\*\z4
%%STOP
\nn \\[0.5mm] & & \mbox{\hspp}
%%START
+\frac{412510829265}{1075095164}\,\*\z5\Big)
%%STOP
\nn \\[0.5mm] & & \mbox{\hspp}
%%START
+\nf\,\*\cfs\,\*\ca\,\*\Big(\frac{263495977106944686412539673552524792748083140531536877569143977}{419116399458783783033438292353667627586148116282566656000000}
%%STOP
\nn \\[0.5mm] & & \mbox{\hspp}
%%START
+\frac{426838279565664147292045705720627}{2427120482554862755056972576000}\,\*
\z3
-\frac{1148280565216821781729}{10073534757714988560}\,\*\z4
%%STOP
\nn \\[0.5mm] & & \mbox{\hspp}
%%START
-\frac{6908656154875}{11288499222}\,\*\z5\Big)
%%STOP
\nn \\[0.5mm] & & \mbox{\hspp}
%%START
+\nf\,\*\cft\,\*\Big(-\frac{360559931378243878964783136466534067956903437618887843864699}{1319532150991841898570447201428312089999679232688128000000}
%%STOP
\nn \\[0.5mm] & & \mbox{\hspp}
%%START
-\frac{10661216839059385141837716718993}{82742743723461230286033156000}\,\*\z3
+\frac{578183477935041694523}{30220604273144965680}\,\*\z4
%%STOP
\nn \\[0.5mm] & & \mbox{\hspp}
%%START
+\frac{77527598620}{245402157}\,\*\z5\Big)
+\nfs\,\*\dfRRna\,\*\Big(-\frac{2383933818125968184095797251}{46488224159447278336128000}
%%STOP
\nn \\[0.5mm] & & \mbox{\hspp}
%%START
-\frac{74838165463102148}{6145124353087005}\,\*\z3
+\frac{40643840}{576081}\,\*\z5\Big)
%%STOP
\nn \\[0.5mm] & & \mbox{\hspp}
%%START
+\nfs\,\*\cas\,\*\Big(-\frac{6462650086648019415443026598069441636721506100243643}{939331677114521371824743418186488047749235814400000}
%%STOP
\nn \\[0.5mm] & & \mbox{\hspp}
%%START
-\frac{354099245860762766307319}{13762463185990217370672}\,\*\z3
+\frac{7261415246443}{611997922107}\,\*\z4
+\frac{2540240}{1728243}\,\*\z5\Big)
%%STOP
\nn \\[0.5mm] & & \mbox{\hspp}
%%START
+\nfs\,\*\cf\,\*\ca\,\*\Big(-\frac{127782475417386997739503642627913187626862415117794169}{316867885746631876095546779734908634774075548057600000}
%%STOP
\nn \\[0.5mm] & & \mbox{\hspp}
%%START
+\frac{1229037023912178380313613}{31759530429208193932320}\,\*\z3
-\frac{152491267750345}{8567970909498}\,\*\z4\Big)
%%STOP
\nn \\[0.5mm] & & \mbox{\hspp}
%%START
+\nfs\,\*\cfs\,\*\Big(\frac{7435370819699556380183158314263420682408867613108511183}{950603657239895628286640339204725904322226644172800000}
%%STOP
\nn \\[0.5mm] & & \mbox{\hspp}
%%START
-\frac{1281373405931715198600583}{103218473894926630280040}\,\*\z3
+\frac{200914839131}{33865497666}\,\*\z4\Big)
%%STOP
\nn \\[0.5mm] & & \mbox{\hspp}
%%START
+\nft\,\*\ca\,\*\Big(\frac{126343754127232276875953655744089}{785539349741617315822338641280000}
-\frac{167133439037}{152394739497}\,\*\z3\Big)
%%STOP
\nn \\[0.5mm] & & \mbox{\hspp}
%%START
+\nft\,\*\cf\,\*\Big(\frac{1384185768719724959140935874368044957253277}{66578470860020623662149109920338746168960000}
+\frac{33786002140459}{38555869092741}\,\*\z3\Big)
%%;
%%STOP
\; , \\[3mm]
\label{eq:ggq3N22} 
 &&\gamma_{\,\rm gq}^{\,(3)}(N\!=\!22)\:=
%%START
%%L texggq3N22=
\dfRAnc\,\*\Big(-\frac{3810637698789182589166962695084388793}{89230400485397403301745762486400000}
%%STOP
\nn \\[0.5mm] & & \mbox{\hspp}
%%START
-\frac{6094498287658989563858609839}{40146132745472682621452760}\,\*\z3
+\frac{8917710298240}{39509747277}\,\*\z5\Big)
%%STOP
\nn \\[0.5mm] & & \mbox{\hspp}
%%START
+\cf\,\*\cat\,\*\Big(\frac{838168957919971745941882717701976959605144167071066891224479}{65195884360255255138534845477237186513400818088399257600000}
%%STOP
\nn \\[0.5mm] & & \mbox{\hspp}
%%START
-\frac{63925515097477464908665329999181}{2548476506682605892809821204800}\,\*
\z3
-\frac{238991866849751167}{1175245721733415332}\,\*\z4
%%STOP
\nn \\[0.5mm] & & \mbox{\hspp}
%%START
-\frac{156381446930}{6238381149}\,\*\z5\Big)
%%STOP
\nn \\[0.5mm] & & \mbox{\hspp}
%%START
+\cfs\,\*\cas\,\*\Big(-\frac{7033360499414036351789237883502252039714921571789277061765421}{26194774966173986439589893272104226724134257267660416000000}
%%STOP
\nn \\[0.5mm] & & \mbox{\hspp}
%%START
-\frac{10304281919862409454000321308469}{65012155782719538081883194000}\,\*\z3
+\frac{1551809103986783059567}{52886057478003689940}\,\*\z4
%%STOP
\nn \\[0.5mm] & & \mbox{\hspp}
%%START
+\frac{5782041496870}{13169915759}\,\*\z5\Big)
%%STOP
\nn \\[0.5mm] & & \mbox{\hspp}
%%START
+\cft\,\*\ca\,\*\Big(\frac{206003883349125053689977262040591693421768120833271039002381393}{419116399458783783033438292353667627586148116282566656000000}
%%STOP
\nn \\[0.5mm] & & \mbox{\hspp}
%%START
+\frac{47163103448793652635305347261}{103133761763573470798118250}\,\*\z3
-\frac{381678739491889796758}{4407171456500307495}\,\*\z4
%%STOP
\nn \\[0.5mm] & & \mbox{\hspp}
%%START
-\frac{1530801120040}{1717815099}\,\*\z5\Big)
%%STOP
\nn \\[0.5mm] & & \mbox{\hspp}
%%START
+\cff\,\*\Big(-\frac{115440871244206858481225194035838370454078329010606897935524799}{488969132701914413539011341079278898850506135662994432000000}
%%STOP
\nn \\[0.5mm] & & \mbox{\hspp}
%%START
-\frac{66241113405753213469306855164743}{227542545239518383286591179000}\,\*\z3
+\frac{342044082410941}{5952229318035}\,\*\z4
+\frac{2799408517960}{5644249611}\,\*\z5\Big)
%%STOP
\nn \\[0.5mm] & & \mbox{\hspp}
%%START
+\nf\,\*\dfRRnc\,\*\Big(-\frac{2383933818125968184095797251}{81354392279032737088224000}
-\frac{299352661852408592}{43015870471609035}\,\*\z3
%%STOP
\nn \\[0.5mm] & & \mbox{\hspp}
%%START
+\frac{162575360}{4032567}\,\*\z5\Big)
%%STOP
\nn \\[0.5mm] & & \mbox{\hspp}
%%START
+\nf\,\*\cf\,\*\cas\,\*\Big(-\frac{107540338447339584673051150618836693843786032479407041}{36164269568909072815252621600179789838345578854400000}
%%STOP
\nn \\[0.5mm] & & \mbox{\hspp}
%%START
+\frac{7193334729924510201427}{246934148073987153780}\,\*\z3
-\frac{368842447989692}{29987898183243}\,\*\z4
-\frac{133906880}{12097701}\,\*\z5\Big)
%%STOP
\nn \\[0.5mm] & & \mbox{\hspp}
%%START
+\nf\,\*\cfs\,\*\ca\,\*\Big(\frac{249167231095782367164545318724726003893497926781218979}{831778200084908674750810296804135166281948313651200000}
%%STOP
\nn \\[0.5mm] & & \mbox{\hspp}
%%START
-\frac{1927955699252891935475921}{180632329316121602990070}\,\*\z3
+\frac{637357131311632}{29987898183243}\,\*\z4
-\frac{40640}{15939}\,\*\z5\Big)
%%STOP
\nn \\[0.5mm] & & \mbox{\hspp}
%%START
+\nf\,\*\cft\,\*\Big(-\frac{9106795540729822447252515307533949880691219005898049}{24464064708379666904435596964827504890645538636800000}
%%STOP
\nn \\[0.5mm] & & \mbox{\hspp}
%%START
-\frac{476362175359316632635209}{25804618473731657570010}\,\*\z3
-\frac{1061322858980}{118529241831}\,\*\z4
+\frac{81280}{5313}\,\*\z5\Big)
%%STOP
\nn \\[0.5mm] & & \mbox{\hspp}
%%START
+\nfs\,\*\cf\,\*\ca\,\*\Big(-\frac{37657295064789710217102953122806488241419}{29239556811603260282015419376521188480000}
-\frac{1024894785160}{1066763176479}\,\*\z3
%%STOP
\nn \\[0.5mm] & & \mbox{\hspp}
%%START
+\frac{32512}{15939}\,\*\z4\Big)
+\nfs\,\*\cfs\,\*\Big(\frac{14419170169389405752299622750480576874859567}{11651232400503609140876094236059280579568000}
%%STOP
\nn \\[0.5mm] & & \mbox{\hspp}
%%START
+\frac{2463044656496}{1066763176479}\,\*\z3
-\frac{32512}{15939}\,\*\z4\Big)
+\nft\,\*\cf\,\*\Big(\frac{1592421949768461095562551154551}{3822714760023908839214731807200}
%%STOP
\nn \\[0.5mm] & & \mbox{\hspp}
%%START
-\frac{32512}{143451}\,\*\z3\Big)
%%;
%%STOP
\; , \\[3mm]
\label{eq:ggg3N22}
 &&\gamma_{\,\rm gg}^{\,(3)}(N\!=\!22)\:=
%%START
%%L texggg3N22=
\dfAAna\,\*\Big(\frac{2059376634376930673356585999995238278647}{611865603328439336926256657049600000}
%%STOP
\nn \\[0.5mm] & & \mbox{\hspp}
%%START
+\frac{4325002786990066966886573851}{417102677875040858404704}\,\*\z3
-\frac{883285615576632050}{55300476272041}\,\*\z5\Big)
%%STOP
\nn \\[0.5mm] & & \mbox{\hspp}
%%START
+\caf\,\*\Big(\frac{145267694114575314964209691288365616920128141866191324687998439229}{109503007371484726530683126463567578467908014061275393064960000}
%%STOP
\nn \\[0.5mm] & & \mbox{\hspp}
%%START
+\frac{48489400214101969658586470078095358711}{42804211406241048575633756955820800}\,\*\z3
-\frac{441642807788316025}{331802857632246}\,\*\z5\Big)
%%STOP
\nn \\[0.5mm] & & \mbox{\hspp}
%%START
+\nf\,\*\dfRAna\,\*\Big(-\frac{2697879353441085526088818397416652567}{19120800104013729278945520532800000}
%%STOP
\nn \\[0.5mm] & & \mbox{\hspp}
%%START
-\frac{6784846923764205269547322444}{3584476137988632376915425}\,\*\z3
+\frac{30896200711520}{16932748833}\,\*\z5\Big)
%%STOP
\nn \\[0.5mm] & & \mbox{\hspp}
%%START
+\nf\,\*\cat\,\*\Big(-\frac{16893932077839621051245851879018529998506918628413933647}{24109513045939381876835081066786526558897052569600000}
%%STOP
\nn \\[0.5mm] & & \mbox{\hspp}
%%START
-\frac{18360340758721692090712631839381}{14002618168585746663790226400}\,\*\z3
+\frac{47564257171}{73301943}\,\*\z4
+\frac{3687397723810}{7256892357}\,\*\z5\Big)
%%STOP
\nn \\[0.5mm] & & \mbox{\hspp}
%%START
+\nf\,\*\cf\,\*\cas\,\*\Big(-\frac{1584314324827178589386027368060768556062335393032264891861}{3327112800339634699003241187216540665127793254604800000}
%%STOP
\nn \\[0.5mm] & & \mbox{\hspp}
%%START
+\frac{108876259075792169315828716211}{224733378014339143986756720}\,\*\z3
-\frac{19495518820936226}{29987898183243}\,\*\z4
+\frac{78487621270}{105172353}\,\*\z5\Big)
%%STOP
\nn \\[0.5mm] & & \mbox{\hspp}
%%START
+\nf\,\*\cfs\,\*\ca\,\*\Big(\frac{6973043222003338465999170594654722169458510456938898093}{23765091430997390707166008480118147608055666104320000}
%%STOP
\nn \\[0.5mm] & & \mbox{\hspp}
%%START
+\frac{12694135944589860757426043}{17840230055913244739760}\,\*\z3
+\frac{53858879220812}{29987898183243}\,\*\z4
-\frac{6652935335300}{5644249611}\,\*\z5\Big)
%%STOP
\nn \\[0.5mm] & & \mbox{\hspp}
%%START
+\nf\,\*\cft\,\*\Big(-\frac{68978030545580684485470547512991272441654878965846967633}{3327112800339634699003241187216540665127793254604800000}
%%STOP
\nn \\[0.5mm] & & \mbox{\hspp}
%%START
-\frac{1399869776455938702809}{1263163142070780440490}\,\*\z3
-\frac{5641743732619}{9995966061081}\,\*\z4
+\frac{123680}{366597}\,\*\z5\Big)
%%STOP
\nn \\[0.5mm] & & \mbox{\hspp}
%%START
+\nfs\,\*\dfRRna\,\*\Big(\frac{1377959702142655845312304584743174897}{728411432533856353483638877440000}
%%STOP
\nn \\[0.5mm] & & \mbox{\hspp}
%%START
+\frac{323429307666373082362}{68319323690202585}\,\*\z3
-\frac{4282807040}{576081}\,\*\z5\Big)
%%STOP
\nn \\[0.5mm] & & \mbox{\hspp}
%%START
+\nfs\,\*\cas\,\*\Big(\frac{7954018871766279472266430722586787152468645331}{216138803951371300004657980031244625244160000}
%%STOP
\nn \\[0.5mm] & & \mbox{\hspp}
%%START
+\frac{95975856624108125795821}{263985866738942788440}\,\*\z3
-\frac{95128514342}{806321373}\,\*\z4
-\frac{267675440}{1728243}\,\*\z5\Big)
%%STOP
\nn \\[0.5mm] & & \mbox{\hspp}
%%START
+\nfs\,\*\cf\,\*\ca\,\*\Big(\frac{86705667957954091736399715900072577365143200631}{932098592040288731270087538884742446365440000}
%%STOP
\nn \\[0.5mm] & & \mbox{\hspp}
%%START
-\frac{365567908653113}{1303821660141}\,\*\z3
+\frac{2000395029470}{16932748833}\,\*\z4\Big)
%%STOP
\nn \\[0.5mm] & & \mbox{\hspp}
%%START
+\nfs\,\*\cfs\,\*\Big(-\frac{1912642678912139708205103378219923049637778657829}{157213962524128699340888098225226559286970880000}
%%STOP
\nn \\[0.5mm] & & \mbox{\hspp}
%%START
+\frac{5312912787577646}{269891083649187}\,\*\z3
-\frac{642112}{4032567}\,\*\z4\Big)
+\nft\,\*\ca\,\*\Big(-\frac{211929916970091160044125}{144505863452897282392056}
%%STOP
\nn \\[0.5mm] & & \mbox{\hspp}
%%START
+\frac{5264045324}{602350749}\,\*\z3\Big)
+\nft\,\*\cf\,\*\Big(-\frac{13090832063043878046580105182292199}{23211524022865174471711851533318400}
-\frac{258064}{12097701}\,\*\z3\Big)
%%;
%%STOP
\; . \nn \\
\eea
}

\def\col{\color{black}}

% ---------------------------------------------------------------------
%
\renewcommand{\theequation}{\ref{sec:appB}.\arabic{equation}}
\setcounter{equation}{0}
\section{Extension and update of the $\bm x$-space approximations}
\label{sec:appB}
%
% ---------------------------------------------------------------------

\vspace{-2mm}
As mentioned above,
we have extended our approximate expressions for $P_{\rm ps}^{\,(3)}(x)$ 
and $P_{\rm qg\,,gq,\,gg}^{\,(3)}(x)$ to $\nf \!=\! 6$. 
The error bands were obtained in the same manner as those for 
$\nf = 3,\,4,\,5$ in refs.~
\cite{Falcioni:2023luc,Falcioni:2023vqq,Falcioni:2024xyt,Falcioni:2024qpd}.
The chosen approximations for $P_{\rm ps\,,qg,\,gg}^{\,(3)}(\nf,x)$ 
are given by
\bea
\label{eq:Pps3A3-nf6}
{\lefteqn{
 P_{\rm ps,\,A}^{\,(3)}(6,x) \; = \;
 p_{{\rm ps},0}^{\,(\nf=6)}(x) 
     + 134701\,\*x_1\*L_0/x
     + 518318\,\*x_1/x
     - 195241\,\*x_1\*(1\!+\!2\*x)
     + 66517\,\*x_1\*x^2
}}
\nn \\ && \mbox{}
     + 658832\,\*x_1\*L_0
     + 19605\,\*L_0^2
     + 76125\,\*L_0^3
     - 4734.5\,\*x_1\*L_1
     - 2035.2\,\*x_1\*L_1^2
     + 1633.1\,\*x_1^2\*L_1^2
\, ,
\nn\\[1mm]
{\lefteqn{
 P_{\rm ps,\,B}^{\,(3)}(6,x) \; = \;
 p_{{\rm ps},0}^{\,(\nf=6)}(x) 
     + 110032\,\*x_1\*L_0/x
     + 341158\,\*x_1/x
     - 365676\,\*x_1
     + 25934\,\*x_1\*x\*(1+x)
}}
\nn \\ && \mbox{}
     + 3614.4\,\*x_1\*L_0
     - 194868\,\*L_0^2
     - 4172.2\,\*L_0^3
     + 3924.3\,\*x_1\*L_1
     - 1324.9\,\*x_1\*L_1^2
     - 12520\,\*x_1^2\*L_1^2
\, ,
\\[4mm]
{\lefteqn{
 P_{\rm qg,\,A}^{\,(3)}(6,x) \; = \;
 p_{{\rm qg},0}^{\,(\nf=6)}(x) 
     + 375000\,\*L_0/x
     + 1595330\,\*x_1/x
     - 477729
     + 637552\,x\*(2-x)
}}
\nn \\ && \mbox{}
     + 931556\,\*L_0
     - 387017\,\*L_0^2
     + 187509\,\*L_0^3
     + 91373\,\*L_1
     + 20710\,\*L_1^2
     +  715.5\,\*L_1^3
     - 346374\,\*L_0\*L_1
\, ,
\nn\\[1mm]
{\lefteqn{
 P_{\rm qg,\,B}^{\,(3)}(6,x) \; = \;
 p_{{\rm qg},0}^{\,(\nf=6)}(x) 
     + 270000\,\*L_0/x
     + 912695\,\*x_1/x
     - 200034
     - 189918\,x\*(2-x)
}}
\nn \\ && \mbox{}
     + 603114\,\*L_0
     - 190521\,\*L_0^2
     + 56661\,\*L_0^3
     - 150856\,\*L_1
     - 16453\,\*L_1^2
     - 1224.3\,\*L_1^3
     + 410661\,\*L_0\*L_1
\, ,
\nn\\
\\[-1mm]
\label{eq:Pgg3A3-nf6}
{\lefteqn{
 P_{\rm gg,\,A}^{\,(3)}(6,x) \; = \;
 p_{{\rm gg},0}^{\,[\nf=6]}(x) 
      - 476018 \,\* x_1\*L_0/x
      - 469289 \,\* x_1/x
      + 2049351 \,\* x_1
      - 1589000 \,\* x_1\*x
}}
\nn \\ && \mbox{}
      + 3185549 \,\* x_1\*L_0
      + 1994521 \,\* L_0^{2}
      + 527723 \,\* L_0^{3}
      - 340674 \,\* x_1\*L_1
      +  22460 \,\* x_1\*L_1^2
      - 394556 \,\* L_0\*L_1
\, ,
\nn\\[1mm]
{\lefteqn{
 P_{\rm gg,\,B}^{\,(3)}(6,x) \; = \;
 p_{{\rm gg},0}^{\,[\nf=6]}(x)
            - 709863 \,\*  x_1\*L_0/x
            - 2134347 \,\* x_1/x
            + 1605315 \,\* x_1\*x
            + 360743 \* x_1\*(2-x^{2})
}}
\nn \\ && \mbox{}
            - 2426250 \,\* x_1\*L_0
            + 230631 \,\* L_0^{2}
            - 185804 \,\* L_0^{3}
            - 7992.9 \,\* x_1\*L_1
            + 15918 \,\* x_1\*L_1^2
            - 32771 \,\* x_1^{2}\*L_1
\eea
in terms of $x_1^{} = 1\!-\!x$, $L_1=\ln \xm1$, $L_0 = \ln x$,
where $p_{{\rm ps},0}^{\,(\nf)}(x)$ has been given in eq.~(26) of
ref.~\cite{Falcioni:2023luc}, 
$p_{{\rm qg},0}^{\,(\nf)}(x)$ in eq.~(15) of ref.~\cite{Falcioni:2023vqq},
and $p_{{\rm gg},0}^{\,[\nf]}(x)$ by eqs.~(10), (11) and (13) of
ref.~\cite{Falcioni:2024qpd}.

Note that the coefficient 142612 of $L_0^3$ in the second line of eq.~(18)
in ref.~\cite{Falcioni:2024xyt} was misprinted as 42612%
\footnote{We thank Thomas Cridge for pointing this out.}; 
the {\sc Fortran} file distributed with that paper is, of course, correct. 

The corresponding approximations for $P_{\rm gq}^{\,(3)}(x)$ 
at $\nf = 3,\,4,\,5$ in ref.~\cite{Falcioni:2024xyt} include an 
estimate of the size of the coefficient of $x^{-1} \ln x$ relative 
to that of the $x^{-1} \ln^2 x$ contribution (denoted by $\bfkl1$) 
as obtained via the gluon-gluon case in ref.~\cite{Moch:2023tdj}. 
The results of ref.~\cite{Falcioni:2024qpd}, as well as the $N=22$ moment
for the quark-to-gluon case reported in the present paper, suggest a 
moderate but not insignificant shift of the $x^{-1} \ln x$ coefficient, 
from $(3 \ldots 6)\, \bfkl1$ to $(3.5 \ldots 7)\, \bfkl1$. 

\pagebreak

We have taken this improved estimate into account not only for the new
$P_{\rm gq}^{\,(3)}(\nf,x)$ at $\nf=6$ but also for $\nf = 3,\,4,\,5$,
i.e., we update the approximations provided in ref.~\cite{Falcioni:2024xyt}.
Our present best estimates for the error bands of this splitting function
read
\bea
\label{eq:Pgq3A3-nf3}
{\lefteqn{
 P_{\rm gq,\,A}^{\,(3)}(3,x) \; = \;
 p_{{\rm gq},0}^{\,(\nf=3)}(x) +
             3.5\,\*\bfkl1\,\*L_0/x
           - 27891 \,\*x_1/x
           - 309124
           + 1056866\,x\*(2-x)
}}
\nn \\ && \mbox{}
           - 124735 \,\*L_0
           - 16246 \,\*L_0^2
           + 131175 \,\*L_0^3
           + 343181 \,\*L_1
           + 60041 \,\*L_1^2
           + 4970.1 \,\*L_1^3
           - 958330 \,\*L_0\*L_1
\, ,
\nn\\[1mm]
{\lefteqn{
 P_{\rm gq,\,B}^{\,(3)}(3,x) \; = \;
 p_{{\rm gq},0}^{\,(\nf=3)}(x) +
             7\,\bfkl1\,\*L_0/x
           - 1139334\,\*x_1/x
           + 143008
           - 290390\,x\*(2-x)
}}
\nn \\ && \mbox{}
           - 659492\,\*L_0
           + 303685\,\*L_0^2
           - 81867 \,\*L_0^3
           - 51206 \,\*L_1
           - 465.9 \*L_1^2
           + 1811.8 \*L_1^3
           + 274249 \,\*L_0\*L_1
,
\nn \\
\\[-2mm]
\label{eq:Pgq3A3-nf4}
{\lefteqn{
 P_{\rm gq,\,A}^{\,(3)}(4,x) \; = \;
 p_{{\rm gq},0}^{\,(\nf=4)}(x) +
             3.5\,\*\bfkl1\,\*L_0/x
           - 8302.8 \,\*x_1/x
           - 347706
           + 1105306\,x\*(2-x)
}}
\nn \\ && \mbox{}
           - 127650 \,\*L_0
           - 29728 \,\*L_0^2
           + 137537 \,\*L_0^3
           + 345513 \,\*L_1
           + 59205 \,\*L_1^2
           + 4658.1 \,\*L_1^3
           - 995120\,\*L_0\*L_1
\, ,
\nn\\[1mm]
{\lefteqn{
 P_{\rm gq,\,B}^{\,(3)}(4,x) \; = \;
 p_{{\rm gq},0}^{\,(\nf=4)}(x) +
             7\,\bfkl1\,\*L_0/x
           - 1129822 \,\*x_1/x
           + 108527
           - 254166\,x\*(2-x)
}}
\nn \\ && \mbox{}
           - 667254\,\*L_0
           + 293099\,\*L_0^2
           - 77437  \,\*L_0^3
           - 52451 \,\*L_1
           - 1850.3  \,\*L_1^2
           + 1471.3 \,\*L_1^3
           + 248634 \,\*L_0\*L_1
\, ,
\nn \\
\\[-2mm]
\label{eq:Pgq3A3-nf5}
{\lefteqn{
 P_{\rm gq,\,A}^{\,(3)}(5,x) \; = \;
 p_{{\rm gq},0}^{\,(\nf=5)}(x) +
             3.5\,\*\bfkl1\,\*L_0/x
           + 14035 \,\*x_1/x
           - 384003
           + 1152711\,x\*(2-x)
}}
\nn \\ && \mbox{}
           - 126346  \,\*L_0
           - 42967  \,\*L_0^2
           + 144270 \,\*L_0^3
           + 348988 \,\*L_1
           + 58688 \,\*L_1^2
           + 4385.5 \,\*L_1^3
           - 1031165\,\*L_0\*L_1
\, ,
\nn\\[1mm]
{\lefteqn{
 P_{\rm gq,\,B}^{\,(3)}(5,x) \; = \;
 p_{{\rm gq},0}^{\,(\nf=5)}(x) +
             7\,\bfkl1\,\*L_0/x
           - 1117561 \,\*x_1/x
           + 76329
           - 218973 \,x\*(2-x)
}}
\nn \\ && \mbox{}
           - 670799 \,\*L_0
           + 282763 \,\*L_0^2
           - 72633  \,\*L_0^3
           - 52548 \,\*L_1
           - 2915.5 \,\*L_1^2
           + 1170.0  \,\*L_1^3
           + 223771 \,\*L_0\*L_1
\, ,
\nn \\
\\[-2mm]
\label{eq:Pgq3A3-nf6}
{\lefteqn{
 P_{\rm gq,\,A}^{\,(3)}(6,x) \; = \;
 p_{{\rm gq},0}^{\,(\nf=6)}(x) +
             3.5\,\*\bfkl1\,\*L_0/x
           + 39203 \,\*x_1/x
           - 417914
           + 1199042 \,x\*(2-x)
}}
\nn \\ && \mbox{}
           - 120750 \,\*L_0
           - 55941  \,\*L_0^2
           + 151383 \,\*L_0^3
           + 353589 \,\*L_1
           + 58466 \,\*L_1^2
           + 4149.2 \,\*L_1^3
           - 1066510\,\*L_0\*L_1
\, ,
\nn\\[1mm]
{\lefteqn{
 P_{\rm gq,\,B}^{\,(3)}(6,x) \; = \;
 p_{{\rm gq},0}^{\,(\nf=6)}(x) +
             7\,\bfkl1\,\*L_0/x
           - 1102470 \,\*x_1/x
           + 46517
           - 184858 \,x\*(2-x)
}}
\nn \\ && \mbox{}
           - 670056 \,\*L_0
           + 272689 \,\*L_0^2
           - 67453 \,\*L_0^3
           - 51523 \,\*L_1
           - 3686.2 \,\*L_1^2
           + 905.0 \,\*L_1^3
           + 199594 \,\*L_0\*L_1
\, ,
\nn \\
\eea

\vspace{4mm}

{\sc Form} and {\sc Fortran} files of the results in these appendices
and the all-$N$ expressions for the $\zeta_3$~terms in 
eqs.~(\ref{eq:gps3z3N}), (\ref{eq:gqg3z3N}) and (\ref{eq:ggq3z3N}) 
have been deposited at the preprint server {\tt http://arXiv.org}. 
These~files are also available from the authors upon request.

\end{document}